\documentclass[12pt]{article}
\usepackage{natbib}

\usepackage{amsfonts}
\usepackage{amssymb}
\usepackage{epsfig}

\usepackage{amsmath}
\usepackage{amsbsy}

\def \QED{\hfill{\Box}\bigskip}

\newtheorem{definition}{Definition}
\newtheorem{property}{Property}

\begin{document}
\title{Nonparametric estimation of galaxy cluster's emissivity and point source detection in astrophysics with two lasso penalties}
\date{\today}
\author{Jairo Diaz Rodriguez, Dominique Eckert, Hatef Monajemi,\\ St\'ephane Paltani  and Sylvain Sardy}
\maketitle

{\bf Abstract}: Astrophysicists are interested in recovering the 3D gas emissivity of a galaxy cluster from a 2D image taken by a telescope.
A blurring phenomenon and presence of point sources make this inverse problem even harder to solve. The current state-of-the-art technique is two step: first identify the location of potential point sources,
then mask these locations and deproject the data.

We instead model the data as a Poisson generalized linear model (involving blurring, Abel and wavelets operators)
regularized by two lasso penalties to induce sparse wavelet representation and sparse point sources.
The amount of sparsity is controlled by two quantile universal thresholds.
As a result, our method outperforms the existing one. 

\newpage

\section{Introduction}

\subsection{Emissivity of astrophysical sources}

Several types of astrophysical sources originate from the radiative processes occurring in an ``optically thin'' environment,
that is, a situation in which a photon has a low probability of interacting with the surrounding material and can escape the source freely.
Such a situation occurs when the mean density of material in the source is very low.
Examples of such astronomical sources include galaxies (where the observed light is the sum of the light emitted by all stars),
the coronae of the Sun and other convective stars, cocoons of expanding material after supernova explosions (\emph{supernova remnants})
and galaxy groups and clusters (which are filled with a hot ($10^7-10^8$ Kelvin) low-density plasma that constitutes the majority
of the ordinary matter of large-scale structures in the Universe).
In case the source is optically thin, the electromagnetic radiation $I$ in a given direction is the integral of the intrinsic emissivity of the source over the source volume,
\begin{equation} I = \frac{1}{4\pi D^2} \int_{V} \varepsilon\,dV, \label{eq:abelint} \end{equation}
where the emissivity $\varepsilon$ is the energy emitted by the source in electromagnetic radiation and $D$ is the source distance.
The three-dimensional distribution of the emissivity is of interest as it provides valuable information on the physical properties of the emitting material (e.g., density, temperature, metallicity).

In the case of galaxy clusters, the emitting plasma is so hot that these structures radiate predominantly in X-rays \citep{sarazin88}.
Current X-ray telescopes like \emph{XMM-Newton} and \emph{Chandra} are able to detect the emission from the plasma and make detailed maps of the distribution of hot gas in galaxy clusters,
which are extremely useful to understand the formation and evolution of structures in the Universe \citep{kravtsov12},
study the overall matter content and the missing mass (``dark matter'') problem \citep{clowe06}, and constrain the cosmological parameters governing  
the evolution of the Universe as a whole \citep{allen11}.
In most cases, X-ray images of galaxy clusters show round, azimuthally symmetric morphologies indicating that the geometry of these structures is nearly spherical.
The observed emissivity decreases radially from the center of the source to its outermost border \citep{e12}.
Assuming spherical symmetry, \eqref{eq:abelint} can be written explicitly as a function of projected distance $s$ to the cluster center,
\begin{equation}
  I(s)\propto\int \varepsilon(r)\, dz \quad \mbox{ with } \quad r^2=s^2+z^2, \label{eq:proj}
\end{equation}
where $r$ is the three-dimensional distance to the cluster center, $I(s)$ is the observed azimuthally-averaged brightness profile, and the integral is performed along the line of sight $z$. 
While $\varepsilon(r)$ can in principle be evaluated directly from the observed emission by solving the integral \eqref{eq:proj},
in practice the problem is rendered complicated by the presence of noise in the original data, as for instance with the XMM-Newton telescope described below.
Indeed, as for all inverse problems the projection kernel smooths small-scale fluctuations, thus the inverse transformation has the opposite effect and the noise can be greatly amplified \citep[see][]{lucy74,lucy94}. This effect is particularly important in the low signal-to-noise regime.


\subsection{The \emph{XMM-Newton} mission}

\begin{figure*}[ht] 
\centering
$
\begin{array}{cc}
\includegraphics[width=2.10in, angle=90]{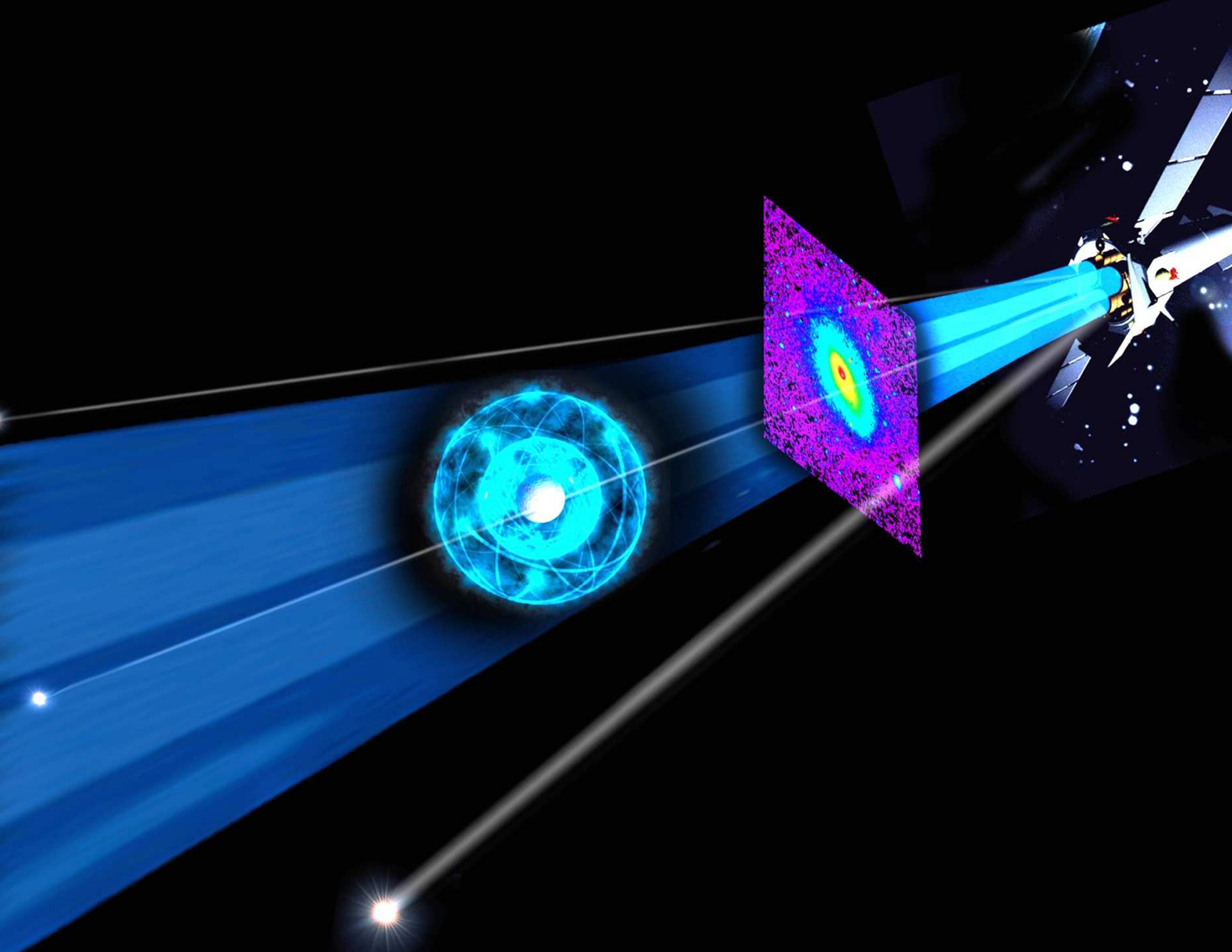}
&
\includegraphics[width=2.28in]{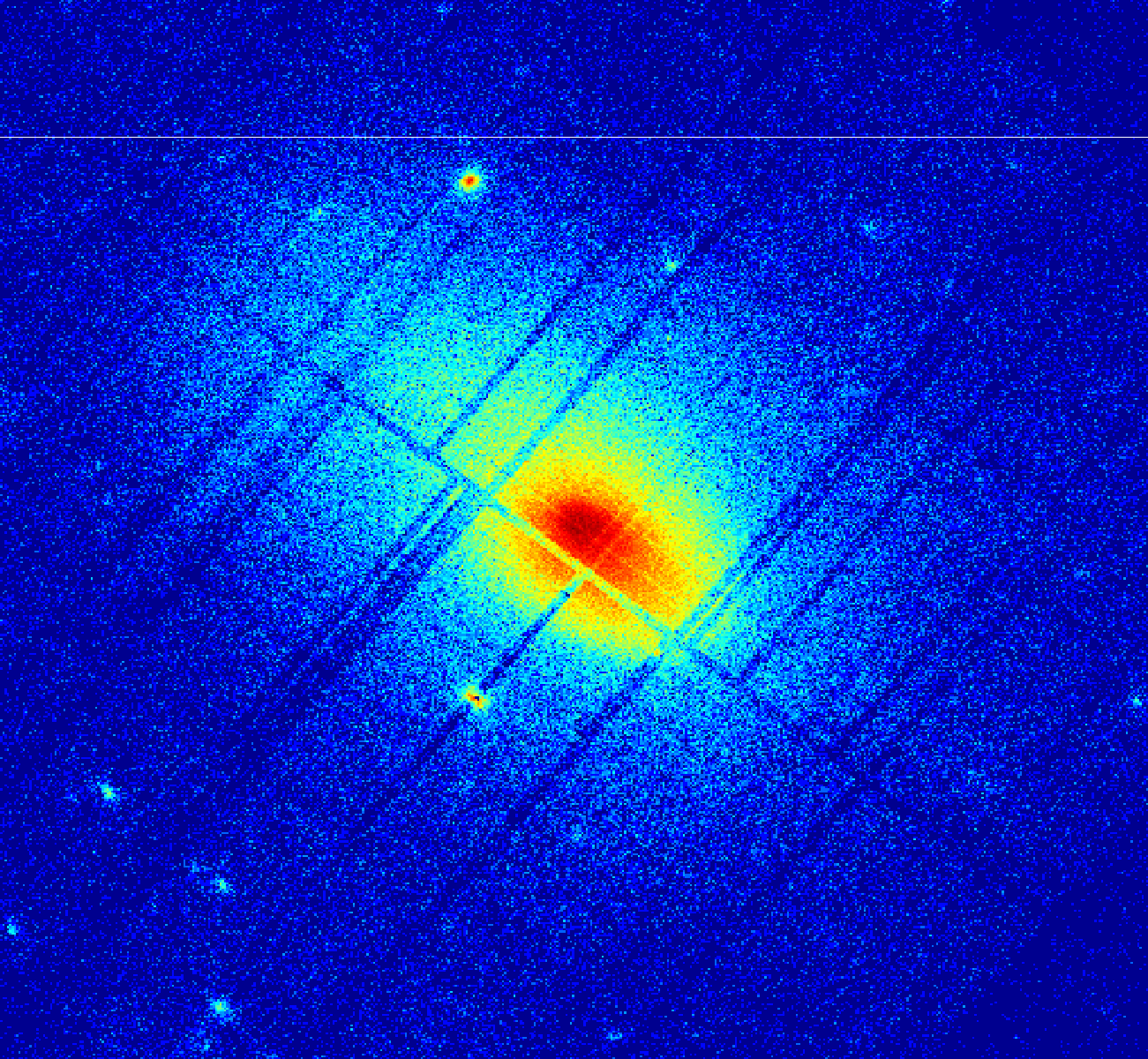}
\end{array}
$
\caption{Left: schematical view of a telescope, the image taken by it, a galaxy cluster and two point sources. Right: real image taken by the XMM-Newton telescope. \label{fig:cosmodata}}
\end{figure*}

The \emph{XMM-Newton} space telescope \citep{jansen01} is a cornerstone mission of the European Space Agency. It was put in orbit on December 10, 1999 by an Ariane 5 launcher and it remains to this day the largest X-ray telescope ever operated. The spacecraft is made of three co-aligned X-ray telescopes that observe the sky simultaneously.
At the focal point of the three telescopes are located two instrument, the European Photon Imaging Camera (EPIC) and the Reflection Grating Spectrometer (RGS). 
 The left image of Figure~\ref{fig:cosmodata} is an image of the galaxy cluster Abell 2142 recorded by the \emph{XMM-Newton} observatory \citep{tchernin16}.
The data were acquired in 2012 (PI: Eckert) as part of the \emph{XMM-Newton} guest observer program, in which astronomers  are invited to propose suitable targets to be observed by the spacecraft and provide a detailed scientific justification for their program.

EPIC \citep{turner01} consists of three high-sensitivity cameras which cover a field of view of 30 arcmin diameter roughly equivalent to the size of the full moon.
The cameras are made of $600\times600$ pixels organized in 8 individual chips which record the time, energy and position of incoming X-ray photons,
resulting in an image like on the right side of Figure~\ref{fig:cosmodata}.
The sensitivity of the instrument is maximal for sources precisely aligned with the axis of the telescopes (the aim point) and gradually declines for sources located slightly offset from the optical axis.
The angular resolution of the telescope is 6 arcsec at the aim point and it degrades to 15 arcsec at the edge of the field of view
Astrophysical sources with an apparent size smaller than the angular resolution of the instrument thus appear blurred with a typical size and shape that is known from the characteristics of the telescopes. Similarly, the degradation of the sensitivity
of the instrument with off-axis angle has been extensively calibrated and follows a known pattern that needs to be taken into account to recover the true flux radiated by a source. 

Apparent on the image of Figure~\ref{fig:cosmodata} are bright spots called point sources.
 The vast majority of these sources are active galactic nuclei, which originate from material falling onto a supermassive black hole located at the center of a galaxy.
 Since they are not originated from the galaxy cluster under study, the estimation of emissivity should be robust to potential point sources.

\subsection{State of the art ``onion peeling'' deprojection} \label{subsct:SA}

Traditionally, the main approach used to solve \eqref{eq:proj} has been by inverting directly the projection kernel \citep[e.g.][]{fabian81,kriss}.
Within the region encompassed between projected radii $r_{i}$ and $r_{i+1}$ from the center of the image of Figure~\ref{fig:cosmodata},
the counts are averaged to give an estimate~$\hat I_i$ of the quantity of radiation received. 
This amounts to discretizing \eqref{eq:proj} such that the projection kernel reduces to an upper-triangular convolution matrix~$V$, where the matrix element $V_{i,j}$ correspond to the volume of the spherical shell $j$ projected along the line of sight of annulus $i$ \citep{kriss}.
The averaged counts $\hat I_i$ are related to the intrinsic 3D emissivity in the spherical shell between $r_{i}$ and $r_{i+1}$  as 
\begin{equation}
\hat I_i=\sum_{j=1}^n V_{i,j} \varepsilon_j + {\rm error} .\label{eq:iter}
\end{equation}
Since the projection matrix $\mathbf{V}$ is upper triangular, the deprojected profile can be evaluated starting from the outermost shell
(where projection effects are assumed to be negligible) and then solving \eqref{eq:iter} iteratively when proceeding inwards (hence the nickname of ``onion peeling''). 

This method has the advantage of being nonparametric in that it makes no assumption on the shape of the intrinsic profile.
It suffers from severe drawbacks however.
As already discussed in the introduction, this method is very sensitive to measurement uncertainties, since small variations in the projected profile can be greatly magnified; therefore, the resulting profile is generally not smooth. Moreover, the propagation of statistical fluctuations can result in unphysical negative emissivities.
This method also requires that the position of contaminating point sources be estimated in a first step, so as to mask the corresponding areas prior to applying the algorithm.

To alleviate these issues, many variants of the direct deprojection technique exist, including a correction for edge effects \citep{mclaughlin99}, spectral information \citep{nulsen95,pizzolato03}, or emission-weighted volumes \citep{morandi07}. However, from the point of view of the mathematical treatment these procedures are similar.

In summary, the current method is a two step method (identify, mask the point sources, and then estimate the emissivity) that does not model well the stochastic nature of the data and that propagates
errors from the outskirt of the galaxy cluster (large radius) to the center of the cluster.

\section{A nonparametric Poisson linear inverse model}

The important stylized features of the astrophysical data described above can be summarized as follows:
\begin{enumerate}
 \item Many bright spots are observed on the image. They are the so-called point sources, that is, sources with an angular size that is much smaller than the angular resolution of the telescope.
 Their location is unknown.
 \item Although point sources are expected to be much smaller than the size of a pixel, their apparent size is much larger.
 This is due to the finite precision of the alignment of the telescope, which induces a blurring effect that has been well studied and can be considered as known.
 \item There are artifacts in the form of lines that are due to the poor sensitivity of the telescope at the connection between the various chips.
 \item Near its center, the image has  a region of high intensity: it is the center of a galaxy cluster where the gas density is high. The emissivity decreases sharply towards the outskirts, implying that the gas density drops radially. The overall shape is nearly spherically symmetric, exception made of the point sources.
 \item Each pixel is a random count of X-rays during a time of exposure.
\end{enumerate}

To account for these specificities, we propose the following model. Considering the telescope first, each image pixel indexed by $(x,y)$ is modeled as
\begin{equation} \label{eq:Poissondata}
  Y_{x,y}\sim{\rm Poisson}(\mu_{x,y}) \quad {\rm for} \quad  x=1,\ldots,N \quad {\rm and} \quad  \ y=1,\ldots,N,
\end{equation}
where $\mu_{x,y}$ reflects the integral of the intrinsic emissivity of the cosmos.
Without the presence of any cosmological background, the XMM telescope has its own electronic noise with small and known mean counts $e_{x,y}\geq 0$.
In other words, without any cosmological object facing the telescope, we have $\mu_{x,y}=e_{x,y}$, which can be seen as a known offset.

Considering now the cosmos, each pixel faces a region of the cosmos along a line going from zero (the captor) to infinity.
Some lines go through the galaxy cluster, some go through a point source, other go through both.
Calling $\epsilon(x,y,z)\geq 0$ the emissivity of the galaxy cluster along that line and $S_{x,y}\geq 0$ a potential point source,
the integral of the cosmos emissivity along that line is
\begin{equation} \label{eq:alongline}
 I_{x,y} =  \int_0^\infty \epsilon(x,y,z) dz + S_{x,y} \quad {\rm for} \quad  x=1,\ldots,N \quad {\rm and} \quad  \ y=1,\ldots,N.
\end{equation}
Moreover, owing to the rare existence of point sources (see first stylized feature), $S$ is a sparse $N\times N$ matrix.

The connection between $\mu_{x,y}$ and $I_{x,y}$ depends on the characteristics of the telescope.
The blurring effect (second stylized feature) is known through the so-called point spread function of the telescope.
Likewise the sensitivity of the telescope (third stylized feature) is known.
As a result, the Poisson intensity in~\eqref{eq:Poissondata} is modeled as
\begin{equation} \label{eq:firstlinearmodel}
 \mu_{x,y}= e_{x,y} + (B(E \circ {\boldsymbol I}))_{x,y},
\end{equation}
where $B$ is the known blurring  operator, 
$E$ is the known $N \times N$ sensitivity matrix,
and $\circ$ is the notation for the Hadamard product between two matrices.

We pause here to make an important remark. The Poisson counts~\eqref{eq:Poissondata} are linked to the unknown  parameters~\eqref{eq:alongline} though a linear model.
This model belongs to the class of nonparametric generalized linear models \citep{NW72}, but as opposed to the classical approach, the link here must be the identity link.
In other words, the canonical link is not appropriate to properly model the physic.

The unknown objects are the gas  emissivity $\epsilon(x,y,z)$ as well as the location and intensities of the point sources $S$.
An assumption is needed to estimate the three-dimensional gas density function because the problem is unidentifiable in its current form.
Indeed, an infinite number of 3D-functions have the same 2D projection, that is, one cannot recover $\epsilon(x,y,z)$ from $\int \epsilon(x,y,z) dz$.
The fourth stylized feature states that a good approximation of the shape of the galaxy cluster is that it is spherical, that is,
$\epsilon(x,y,z)=\epsilon_R(r)$ with $r=\sqrt{x^2+y^2+z^2}$ is radial.
Invariance by rotation makes the problem simpler since the emissivity is known through a univariate function $\epsilon_R(r)$ of
the distance $r$ to the center must be estimated. The association is moreover linear since the integral in~\eqref{eq:alongline} becomes
 $$
 \int_0^\infty \epsilon(x,y,z) dz=(A \epsilon_R) (x,y), 
 $$
 where $A$ is the Abel transform.
 
The final assumption we make is that $\epsilon_R$ has a sparse representation on basis functions $\phi_p$: 
\begin{equation}\label{eq:linearexpansion}
 \epsilon_R(r)=\alpha_0+\sum_{p=1}^P \alpha_p \phi_p(r).
\end{equation}
The choice of basis functions $\phi_p$ is based on prior knowledge.
Cosmologists expect a decreasing function from the center of the galaxy cluster to its outskirt.
So we use a generalization of the so-called \emph{King}'s functions
\begin{equation}
 \phi_p(r)=(1+ (r/\rho)^2)^{-\beta}, \quad \rho \in \{\rho_1, \ldots, \rho_I\},\, \beta \in \{\beta_1, \ldots, \beta_J\}
\end{equation}
parametrized by $p=(\rho,\beta)$ \citep{2016A&A...592A..12E}. A grid of  $(\rho,\beta)$ lead to $P/2$ such functions.
To allow more flexibility and discover galaxy clusters with singularities, we also use $P/2$ orthonormal wavelets defined on equispaced radii.
Here we chose $P$ of the order of $N$, more precisely $P=2^{\lfloor\log_2(N) \rfloor}$.
We provide more details of our implementation in Appendix~\ref{app:waveletimplement}.

Putting all components together leads to the following linear model for the Poisson parameters:
\begin{equation}\label{eq:poismodel}
 \mu_{x,y}=e_{x,y}+ (B(E \circ (A (\alpha_0 {\bf 1}  + \Phi {\boldsymbol \alpha})+{\bf s})))_{x,y},
\end{equation}
where the unknown parameters are the intercept $\alpha_0$, the sparse $N$-vector ${\boldsymbol \alpha}$ of the linear expansion~\eqref{eq:linearexpansion}
and the sparse $N \times N$-matrix $S$ of potential point sources put in vector form ${\bf s}$.
This is a linear inverse problem in the sense that the unknown quantities are indirectly observed through the linear operators.

\section{Estimation with two sparsity constraints}

Based on stylized feature five, the Poisson negative log-likelihood
\begin{eqnarray} \label{eq:MLE}
-l(\alpha_0,{\boldsymbol \alpha}, S; {\bf y})&=&\sum_{(x,y)\in \{1,\ldots,N\}^2} \mu_{x,y} - Y_{x,y} \log \mu_{x,y} 
\end{eqnarray}
is a natural measure of goodness-of-fit of the counts data to the linear model for  $\mu_{x,y}$~\eqref{eq:poismodel}.
This model is a generalized linear model (GLM) for Poisson noise  with identity link.
Note that the log-term in~\eqref{eq:MLE}  prevents the estimated  Poisson intensities from being negative.

The number $1+N+N^2$ of parameters $(\alpha_0,{\boldsymbol \alpha}, {\bf s})$ exceeds the number of observations $N^2$, so that regularization is needed.
Owing to the sparse representation of the univariate gas density on its basis functions and to the rare existence of point sources, we regularize the likelihood
by enforcing sparsity on the estimation of ${\boldsymbol \alpha}$ and ${\bf s}$ with two $\ell_1$ penalties
\begin{equation} \label{eq:l1penalty}
(\hat \alpha_0, \hat {\boldsymbol \alpha}, \hat {\bf s})_{\lambda_1, \lambda_2} = \arg \min_{\alpha_0,{\boldsymbol \alpha}, {\bf s}} -l(\alpha_0,{\boldsymbol \alpha}, {\bf s}; {\bf y}) + \lambda_1 \|{\boldsymbol \alpha}\|_1
+ \lambda_2 \|{\bf s}\|_1
\end{equation}
in the spirit of lasso \citep{Tibs:regr:1996,SAT01} and  glmnet \citep{ParkHastie07}.
We rely on FISTA \citep{beck2009} to solve the high-dimensional and non-differentiable optimization problem for given hyperparameters $(\lambda_1,\lambda_2)$.
It has the advantage
over glmnet to handle the identity link function and positivity constraints on the King's coefficients, and does no require building and storing a very large matrix.

The selection of the regularization parameters $(\lambda_1,\lambda_2)$ is a key issue.
Performing cross validation on a 2D-grid would be computationally intensive and would require segmenting the image into sub-images.
Another approach is the universal threshold of \citet{Dono94b}.
Derived for Gaussian regression, the universal threshold has the property to reproduce the true signal with high probability when the true signal is the constant function.
This  choice of $\lambda$ has remarkable near minimax properties when the function to estimate lives in Besov's spaces \citep{Dono95asym}.

The quantile universal threshold is the extension of the universal threshold to other noise distributions, models and estimators \citep{Giacoetal16}.
We now derive it for~\eqref{eq:l1penalty}.
First we derive the zero-thresholding function for~\eqref{eq:l1penalty}.
The proof is in Appendix~\ref{app:proof}.

\bigskip
\noindent
\begin{property} \label{prop:ztf}
Given an image ${\bf y}$, the smallest $\lambda_1$ and $\lambda_2$ that jointly set $(\hat {\boldsymbol \alpha}, \hat s)_{\lambda_1,\lambda_2}$ in~\eqref{eq:l1penalty} to zero is 
given by the zero-thresholding function
\begin{equation}
  \lambda({\bf y})=(\lambda_1({\bf y}), \lambda_2({\bf y})):= \left \{
   \begin{array}{ll}
   \left ( \|X_1^{\rm T} \left ( \frac{ {\bf y} -  \hat {\boldsymbol \mu}_\lambda(\hat \alpha_0) } {\hat {\boldsymbol \mu}_\lambda(\hat \alpha_0) } \right ) \|_\infty ,
   \|X_2^{\rm T} \left ( \frac{ {\bf y} -  \hat {\boldsymbol \mu}_\lambda(\hat \alpha_0) } {\hat {\boldsymbol \mu}_\lambda(\hat \alpha_0) } \right ) \|_\infty \right )& {\rm if} \ {\bf y} \in {\cal D} \\
  (+ \infty, + \infty) & {\rm otherwise}
   \end{array}
  \right . ,
\end{equation}
where $\hat {\boldsymbol \mu}_\lambda(\hat \alpha_0)={\bf e}+ {\bf x}_0 \hat \alpha_0$, ${\bf x}_0=BE \circ A {\bf 1}$,
$X_1=BE \circ A  \Phi$, $X_2=BE \circ A$ and ${\cal D} = \{ {\bf y} : \exists \hat \alpha_0 \in {\mathbb R} \ {\rm satisfying}
\ \hat {\bf x}_0^{\rm T} {\bf 1}= {\bf x}_0^{\rm T} ({\bf y}/({\bf e}+ {\bf x}_0 \hat \alpha_0)) \ {\rm and} \ {\bf e}+ {\bf x}_0 \hat \alpha_0 > {\bf 0}\}$.
\end{property}

%

\bigskip
\noindent

Second we define the corresponding null-thresholding statistic.

\begin{definition} \label{def:nts}
 The null-thresholding statistic $\Lambda$ for $(\hat {\boldsymbol \alpha}, \hat s)_{\lambda_1,\lambda_2}$ in~\eqref{eq:l1penalty} is
$$
\Lambda=(\Lambda_1, \Lambda_2):=(\lambda_1({\bf Y}_0), \lambda_2({\bf Y}_0)) \quad {\rm with} \quad {\bf Y}_0 \sim {\rm Poisson}({\bf e}+{\bf x_0} \alpha_0).
$$
\end{definition}

Note that ${\bf Y}_0$ has mean ${\bf e}+{\bf x_0} \alpha_0$, that is, the zero-scene assumes zero emissivity (i.e., ${\boldsymbol \alpha}={\bf 0}$) and 
no point source (i.e., ${\bf s}={\bf 0}$). The goal of our selected hyperparameters $(\lambda_1^{\rm QUT}, \lambda_2^{\rm QUT})$ is to reproduce this zero-scene with high probability.
This is achieved with the third step by taking marginal quantiles of the null-thresholding statistic.
\noindent
\begin{definition} \label{def:qut}
The quantile universal thresholds $(\lambda_1^{\rm QUT}, \lambda_2^{\rm QUT})$ are the upper $\alpha_1$-quantile of $\Lambda_1$ for $\lambda_1$
and the upper $\alpha_2$-quantile of $\Lambda_2$ for $\lambda_2$.
\end{definition}

The quantile universal thresholds has the following desired property.

\bigskip
\noindent
{\bf Property}: With $(\lambda_1^{\rm QUT}, \lambda_2^{\rm QUT})$, the estimator~\eqref{eq:l1penalty} reproduces the zero-scene with probability at least $1-\alpha_1-\alpha_2$
since ${\mathbb P}((\hat {\boldsymbol \alpha}, \hat {\bf s})_{\lambda_1^{\rm QUT},\lambda_2^{\rm QUT}}= ({\bf 0}, {\bf 0}) ; {\boldsymbol \alpha}={\bf 0}, {\bf s}={\bf 0}) \geq 1-\alpha_1-\alpha_2$.

\bigskip
In practice, the choice of $\alpha_1$ and $\alpha_2$ can be guided by the following considerations.
Since the former is linked to the estimation of the emissivity function $\epsilon_R$, we choose $\alpha_1=1/\sqrt{\pi \log P}$ as for the universal threshold of \citet{Dono94b} in the Gaussian case.
The latter is linked to the identification of the point sources, so we recommend for instance $\alpha_2=1/N^2$ to control the false discovery rate at level $\alpha_2$ in the weak sense:
with $\alpha_2=1/N^2$,  the average number of falsely detected point sources is one per image when no point sources are present.
%
%
%

\section{Numerical experiments}
\label{sec:tests}

\subsection{Simulated data}

\begin{figure}[!t]
\centering
\includegraphics[width=1.0\textwidth]{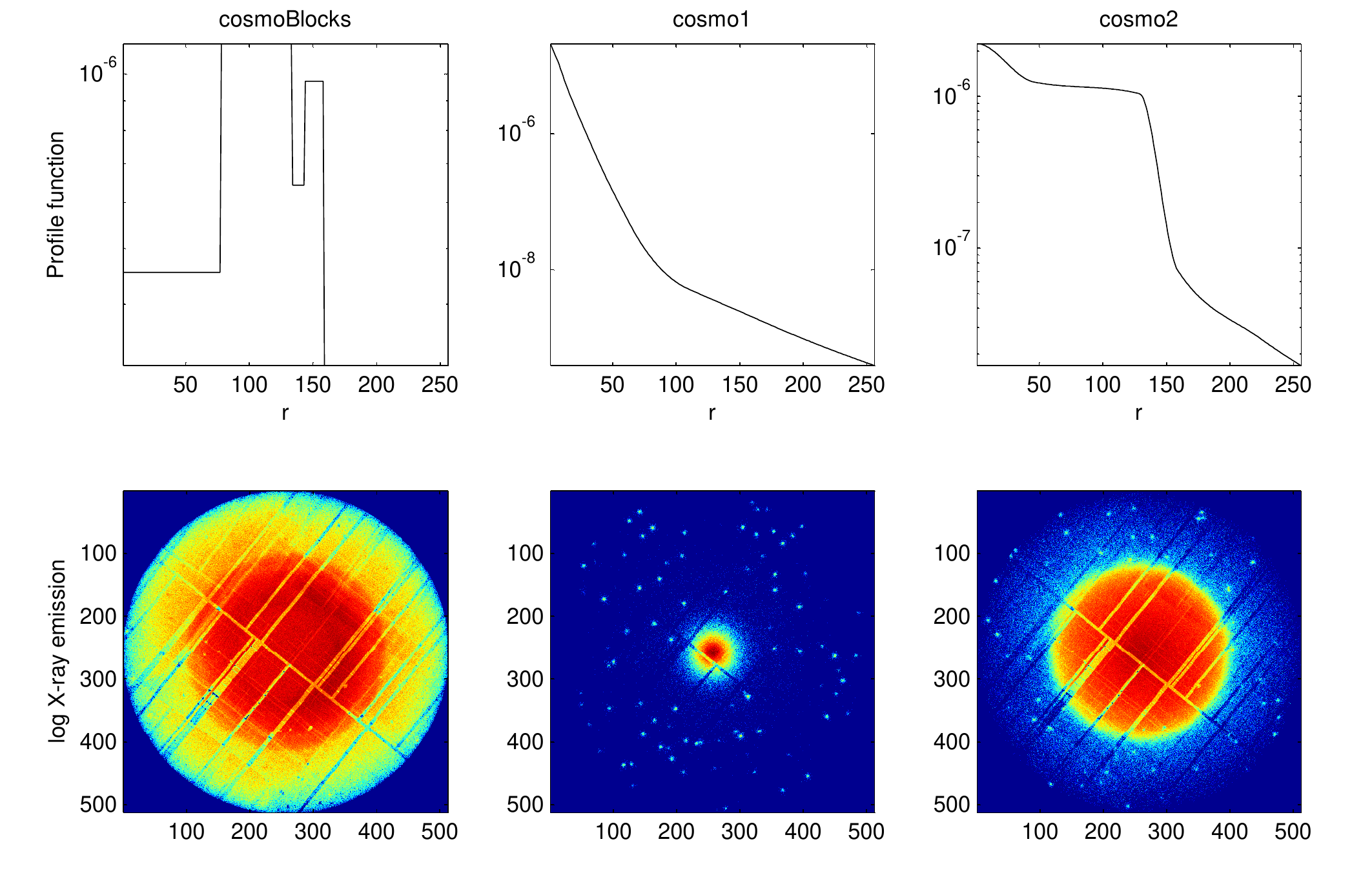}
\caption{Three different simulation profiles (top row) with a corresponding simulated galaxy cluster images (bottom row).}
\label{fig:simprofiles}
\end{figure}

We simulate galaxy clusters according to model~\eqref{eq:poismodel} with known constant background $e_{x,y}=10^{-4}$,
known sensitivity matrix~$E$ and blurring operator~$B$ corresponding to the point spread function
$${\rm psf}(r; r_0, \alpha)=\left(1+\left(\frac{r}{r_0}\right)^2\right)^{-\alpha}$$
 of the XMM telescope ($\alpha=1.449$ and $r_0=2.2364$ pixels).
The simulations are based on three profile functions: 
{\tt cosmoBlocks} is a cropped version of the well known standard function {\tt blocks} used in signal processing \citep{Dono94b}
and although not expected to describe a galaxy cluster, it allows to show the flexibility of our procedure;
{\tt cosmo1} and {\tt cosmo2} are typical profiles according to cosmologists.
For each test profile, we simulate  $N\times N$ images of galaxy clusters for $N \in \{ 128, 256, 512 \}$ en perform $M \in \{96, 48, 24\}$ Monte Carlo samples, respectively, to estimate the mean squared error.  
We consider two scenarios: first without, then with point sources to quantify the robustness of the methods to the presence of point sources.
A total of $N/4$ points sources are uniformly distributed on the whole image.
The amplitude of each point source is uniformly distribution on $[0, 0.002]$.

We compare our estimator (QUT-lasso) to the state-of-the-art method used by cosmologists (SA) described in Section~\ref{subsct:SA}.
Recall that the SA method is a two step method: first estimate the location of potential point sources, then perform the deprojection.
We help the SA method by being oracle in the first step: since we are doing a simulation, we know where the point sources are and provide this information through the sensitivity matrix $E$
in that $E_{x,y}=0$ when pixel $(x,y)$ has a point source.

Table~\ref{meanMSE} shows the estimated mean square error between $\log \hat \epsilon$ and $\log \epsilon$ for each simulation.
The first striking result is that QUT-lasso performs better than the state-of-the-art method, without and with point sources.
Second, as we excepted, QUT-lasso is robust to point sources  by means of the $\ell_1$ penalty on the point source matrix $S$.
The state-of-the-art method is not at all robust for {\tt cosmo1}. 


\begin{table}[ht]
\centering
\caption{Results of Monte-Carlo simulation for images of increasing sizes \emph{without} and \emph{with} point sources. Three tests function plotted on Figure~\ref{fig:simprofiles}
allow to compare two estimators: the proposed QUT-lasso and the state-of-the-art (SA). \label{meanMSE}}
\begin{tabular}{rrrrrrrrrr}
  \hline
  && \multicolumn{8}{c}{MSE of the log-profile (*100)} \\
 \cline{3-10}
 && \multicolumn{2}{c}{\tt cosmoBlocks} && \multicolumn{2}{c}{\tt cosmo1} && \multicolumn{2}{c}{\tt cosmo2} \\
 \cline{3-4} \cline{6-7} \cline{9-10}
 N&&QUT-lasso&SA&&QUT-lasso&SA&&QUT-lasso&SA \\
 \hline
\emph{Without} &  &  &  &  &  &  &  &  &  \\ 
  128 &  & 7.4 & 31 &  & 13 & 37 &  & 30 & 23 \\ 
  256 &  & 3.5 & 15 &  & 2.1 & 17 &  & 1.1 & 13 \\ 
  512 &  & 1.9 & 19 &  & 0.43 & 10 &  & 0.77 & 10 \\ 
   \\
\emph{With} &  &  &  &  &  &  &  &  &  \\ 
  128 &  & 8.6 & 31 &  & 113 & 221 &  & 40 & 16 \\ 
  256 &  & 3.7 & 15 &  & 14 & 66 &  & 2.5 & 10 \\ 
  512 &  & 2.3 & 16 &  & 0.8 & 13 &  & 1 & 11 \\ 
   \hline
\end{tabular}
\end{table}

%

\begin{figure*}[!t]
\centering
\includegraphics[width=0.9\textwidth]{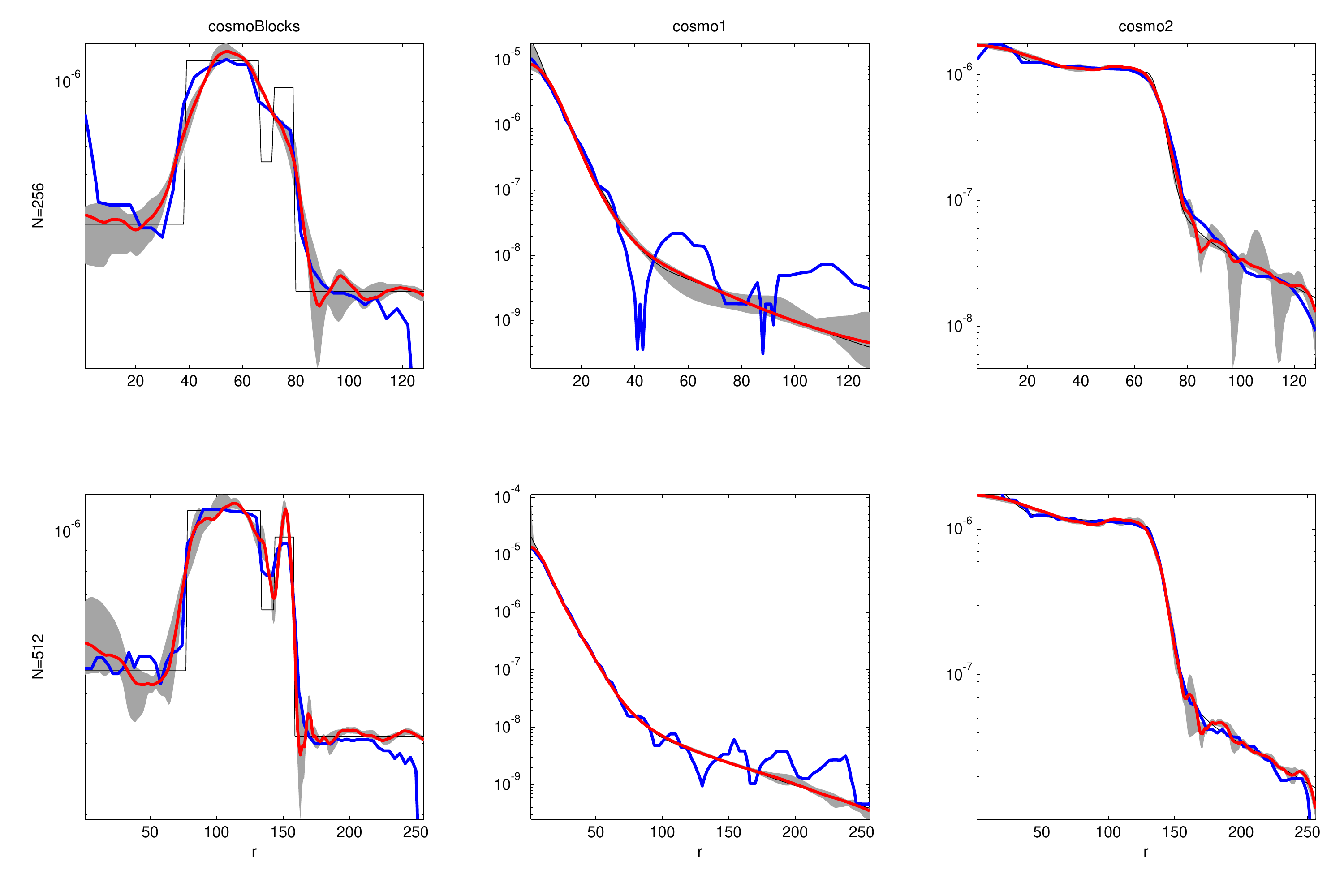}
\caption{Confidence intervals obtained by bootstrap for images of size $N\times N$ with $N=256$ (top) and $N=512$ (bottom).
In black the true profile, in blue the state-of-the-art estimate and in red the estimated profile 
with its confidence intervals in gray obtained with the proposed method.
\label{confidenceintervals}}
\end{figure*}

Cosmologists are also interested in quantifying the uncertainty on the emissivity estimation.
To that aim, the image can be segmented into blocks of size $2 \times 2$ pixels and, assuming
that the four pixels are approximately i.i.d., bootstrapping within each block can be employed to provide bootstrapped images and corresponding emissivity curves. 
Pointwise quantiles of these estimated curves provide a measure of uncertainty, as shown on
Figure \ref{confidenceintervals} for the three test functions and two sample sizes.
We observe that the proposed estimator (red curve) is closer to the true emissivity (black) and less wiggly that the state-of-the-art (red),
and that coverage improves as the sample size increases, especially in areas of discontinuities.

\subsection{Real data}

\begin{figure*}[!t]
\centering
\includegraphics[width=1.0\textwidth, height=6.3in]{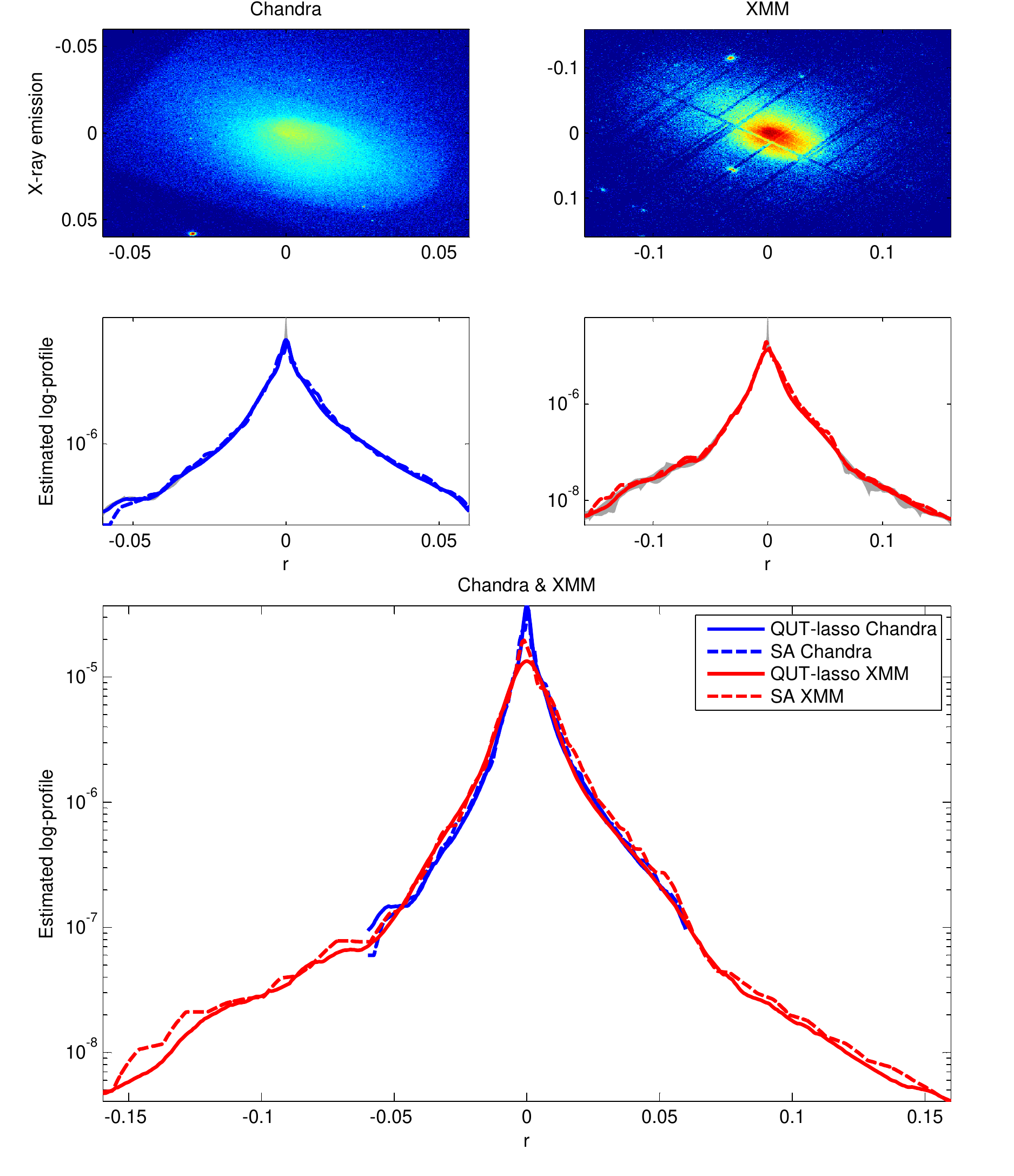}
\caption{Real data results. Top: pictures taken by two telescopes of same galaxy cluster: Chandra (high resolution) and XMM (high sensitivity).
Middle: Estimated emissivities by our method (continuous line) and state-of-the-art (dotted line).
Bottom: all four estimates on the same plot.}
\label{realresult}
\end{figure*}

We applied our method to real data in Figure \ref{realresult}. We obtained confidence intervals by bootstrapping the image in squares of $2 \times 2$.
We also compared to results obtained by the state-of-the-art methodology. The result can be observed in Figure \ref{realresult}. We show the results in the interval $(-0.06,0.06)$ and $(-0.16,0.16)$ for Chandra and XMM-Newton, respectively.

We find an excellent agreement between the results obtained with the two independent telescopes. Given the better spatial resolution of Chandra compared to XMM-Newton, it is able to sample better the shape of the emissivity profile in the innermost regions, whereas in XMM-Newton the peak is smeared out by the point spread function of the telescope. Conversely, the higher sensitivity of XMM-Newton allows it to detect the emission from the source out to larger radii than Chandra.

%
%
%
%

\subsection{Summary of empirical findings}

As shown in Table~1, our method outperforms the current state-of-the-art method by providing results that are typically closer to the true value by a factor of three to five on average.
Thanks to the use of wavelets in the linear expansion~\eqref{eq:linearexpansion}, QUT-lasso  adapts  to local features  of the emissivity.
Moreover our method does not require an \emph{a priori} knowledge of the position of contaminating point sources, but proposes, in a single step, an estimation of the emissivity robust to the presence of point sources.
For the selection of its two regularization parameters, the quantile universal thresholds for Poisson GLM with identity link is employed, which makes the method fully automatic
and superior by far to the methods that are commonly used in astrophysics. 
Figure~\ref{confidenceintervals} also shows good coverage by the bootstrap-based confidence intervals, especially with large sample size.

Application to the Chandra and XMM-Newton telescopes shows  in Figure~\ref{realresult} good agreement  between the profiles reconstructed with QUT-lasso and the standard method,
yet with a smoother profile recovered by QUT-lasso. Given that the true emissivity profile of the source is unknown, we cannot make a quantitative assessment based on this plot. 
However, our results obtained with simulated data clearly highlight the superiority of our method over the current state-of-the-art.

\section{Conclusions}

In this paper, we have presented a novel technique to reconstruct the three-dimensional properties of an ``optically thin'' astrophysical source from two-dimensional observations including the presence of background, unrelated point sources and Poisson noise.
This method is based on Poisson GLM with identity link and a lasso-type regularization with two regularization parameters that are selected with the quantile universal threshold (QUT).
The linear model for the emissivity curve is based on an expansion on basis functions which include wavelets. This makes the QUT-lasso method particularly flexible to discover galaxy clusters
with unusual shapes.

Future applications to real data will allow us to reconstruct accurately the three-dimensional gas density profiles in galaxy clusters, which can be used to study the astrophysical properties of the plasma in clusters of galaxies, estimate cosmological parameters, and measure the gravitational field in massive structures to set constraints on dark matter and modified gravity.

\section{Acknowledgements}

The authors thank the Swiss National Science Foundation.

\section{Reproducible research}

The code and data that generated the figures in this article may be found online at 
{\tt http://www.unige.ch/math/folks/sardy/astroRepository}

\appendix

\section{Proof of Property~\ref{prop:ztf}} \label{app:proof}

The KKT conditions for~\eqref{eq:l1penalty} at ${\boldsymbol \alpha}={\bf 0}$ and ${\bf s}={\bf 0}$ are
\begin{eqnarray*}
\partial/\partial \alpha_0: \ {\bf x}_0^{\rm T} \left ( \frac{{\boldsymbol \mu}-{\bf y}}{{\boldsymbol \mu}}\right ) &=& 0 \\
\nabla_{\boldsymbol \alpha}: \ X_1^{\rm T} \left ( \frac{{\boldsymbol \mu}-{\bf y}}{{\boldsymbol \mu}} \right ) &\in& \lambda_1 {\cal B}^\infty \\
\nabla_{\bf s}: \ X_2^{\rm T} \left (\frac{{\boldsymbol \mu}-{\bf y}}{{\boldsymbol \mu}}\right ) &\in& \lambda_2 {\cal B}^\infty
\end{eqnarray*}
where ${\cal B}^\infty$ is the $\ell_\infty$-unit ball and ${\bf \mu={\bf e}+{\bf x}_0 \alpha_0}$. 
The first equation has a solution provided ${\bf y} \in {\cal D}=\{ {\bf y} : \exists \hat \alpha_0 \in {\mathbb R} \ {\rm satisfying}
\ \hat {\bf x}_0^{\rm T} {\bf 1}= {\bf x}_0^{\rm T} ({\bf y}/({\bf e}+ {\bf x}_0 \hat \alpha_0)) \ {\rm and} \ {\bf e}+ {\bf x}_0 \hat \alpha_0 > {\bf 0}\}$,
and the smallest $\lambda_i$ allowing this system to have a solution are 
$\lambda_i=\|X_i^{\rm T} \left ( \frac{{\boldsymbol \mu}-{\bf y}}{{\boldsymbol \mu}} \right )\|_\infty$ for $i\in \{1,2\}$. $\QED$

\section{Implementation details} \label{app:waveletimplement}

The emissivity function $\epsilon_R(r)$ defined on ${\mathbb R}^+$ typically has a peak at zero and decreases (often monotonically) to zero as $r$ gets large.
Wavelets used in~\eqref{eq:linearexpansion} have difficulties handling such a function because the peak at the left boundary 
is very different from the flat behavior near zero at the right boundary. Various boundary schemes have been proposed. The simplest one
assumes periodicity, which is clearly violated here.
We overcome this difficulty by splitting the original image into two half-images going through the center of the galaxy cluster, for instance
the left image and the right image. Each half faces half of the galaxy cluster. Let us call $\epsilon_R^{{\rm left}}$ and $\epsilon_R^{{\rm right}}$ the corresponding emissivities.
If the galaxy cluster is exactly spherical then $\epsilon_R^{{\rm left}}(r)=\epsilon_R^{{\rm right}}(r)$ for all $r\geq 0$,
otherwise they share the same value at $r=0$ and both tend to zero when the radius $r$ is large.
Hence the double emissivity function $\epsilon_R^{{\rm left}\cup{\rm right}}(r)=\epsilon_R^{{\rm left}}(-r) \cdot  1(r<0) +\epsilon_R^{{\rm right}}(r) \cdot 1(r\geq0)$
defined for negative radii (left part of the galaxy cluster) and for positive radii (right part) can be well represented as a linear combination of wavelets with periodic boundaries.
Plotting both left and right estimated emissivities can reveal asymmetry in the cluster, or can be averaged to provide the cosmologist with a single emissivity curve.
Note that instead of splitting the image into a left and right sectors, one could also split into more sectors where the sphericity assumption seems to better hold.

\bibliographystyle{plainnat}
\bibliography{article}

\end{document}